\documentclass{article}
\usepackage{emulateapj}
\usepackage{psfig}
\usepackage{times}

\newcommand\rozanska{R\'o$\dot{\rm z}$a\'nska }
\newcommand\zycki{$\dot{\rm Z}$ycki }

\newcommand\fx{F_{\rm x}}

\newcommand\dm{\dot{m}}
\newcommand\fdisk{F_{\rm d}}

\newcommand\prad{P_{\rm rad}}

\newcommand\pg{P_{\rm gas}}

\newcommand\pmin{P_{\rm bot}}
\newcommand\pstar{P_*}
\newcommand\tstar{T_*}
\newcommand\thot{\tau_{\rm h}}

\def\>{$>$}
\def\<{$<$}

\def\simlt{\lower.5ex\hbox{$\; \buildrel < \over \sim \;$}}
\def\simgt{\lower.5ex\hbox{$\; \buildrel > \over \sim \;$}}
\def\sqr#1#2{{\vcenter{\hrule height.#2pt
      \hbox{\vrule width.#2pt height#1pt \kern#1pt
         \vrule width.#2pt}
      \hrule height.#2pt}}}

\begin{document}

\title{On the X-ray heated skin of Accretion Disks}
\author{Sergei Nayakshin\altaffilmark{1}}

\affil{NASA/GSFC, LHEA, Code 661, Greenbelt, MD, 20771}
\altaffiltext{1}{National Research Council Associate}

\begin{abstract}
We present a simple analytical formula for the Thomson depth of the
X-ray heated skin of accretion disks valid at any radius and for a
broad range of spectral indices of the incident X-rays, accretion
rates and black hole masses.  We expect that this formula may find
useful applications in studies of geometry of the inner part of
accretion flows around compact objects, and in several other
astrophysically important problems, such as the recently observed
X-ray ``Baldwin'' effect (i.e., monotonic decrease of Fe line's
equivalent width with the X-ray luminosity of AGN), the problem of
missing Lyman edge in AGN, and line and continuum variability studies
in accretion disks around compact objects.  We compute the reflected
X-ray spectra for several representative cases and show that for hard
X-ray spectra and large ionizing fluxes the skin represents a perfect
mirror that does not produce any Fe lines or absorption features. At
the same time, for soft X-ray spectra or small ionizing fluxes, the
skin produces very strong ionized absorption edge and highly ionized
Fe lines that should be observable in the reflected spectra.
\end{abstract}


\section{Introduction}\label{sect:intro}

X-ray illumination of an accretion disk surface is a problem of
general astrophysical interest. Since the X-ray heating of the disk
atmosphere changes energy and ionization balances there, the spectra
emitted by X-ray illuminated accretion disks in any wavelength may be
quite different from those resulting from non-illuminated disks.  It
has been known for many years (e.g., Basko, Sunyaev, \& Titarchuk
1974) that X-ray illumination leads to formation of a hot (and often
completely ionized) X-ray ``skin'' above the illuminated
material. Unfortunately, due to numerical difficulties, most previous
studies of X-ray illumination had to rely on a constant density
assumption for the illuminated gas (e.g., Ross \& Fabian 1993; \zycki
et al. 1994; Matt, Fabian \& Ross 1993; 1996; Ross, Fabian \& Brandt
1996), in which case the completely ionized skin forms only for very
large ionization parameters (e.g., Ross, Fabian \& Young 1999).

Recently, Nayakshin, Kazanas \& Kallman (1999; hereafter paper I) have
shown that if the assumption of the constant density is relaxed, then
the temperature and ionization structure of the illuminated material
is determined by the thermal ionization instability, and that the
X-ray heated skin always forms on the top of the disk (see also
Raymond 1993; Ko \& Kallman 1994; \rozanska \& Czerny 1996). Due to
the presence of the hot skin, the resulting reflected spectra are
quite different from those obtained with the constant density
assumption.

Unfortunately, computations similar to those reported in paper I are
rather numerically involved and time consuming and thus are not
readily performed. In this Letter, we present an approximate
expression for the Thomson depth of the hot skin which allows one to
qualitatively understand effects of the thermal instability on the
reflected spectra. As an example, we apply our approximate expression
to the observed X-ray Baldwin effect (e.g., Nandra et al. 1997b) in
the geometry of a cold accretion disk illuminated by an X-ray source
situated at some height above the black hole.

\section{Thomson Depth of the Skin}\label{sect:approx}

We assume that X-rays are incident on the surface of an accretion disk
whose structure is given by the standard accretion disk theory (e.g.,
SS73). Let us define the hot skin as the region of the gas where
temperature significantly exceeds the effective temperature (see,
e.g., Fig. 3b below). As explained in paper I, even Fe atoms are
completely ionized in the X-ray skin if the X-ray flux, $\fx$, is
comparable with or larger than the disk intrinsic flux, $F_d$, and if
the illuminating X-rays have relatively hard spectra, i.e., photon
index $\Gamma\simlt 2$. Under those conditions, the Compton
temperature, $T_c$, is close to $\sim$ few keV.  Therefore, we can
neglect all the atomic processes and only consider Compton scattering
and bremsstrahlung emission in our analytical study of the X-ray skin.

The Thomson depth of the hot layer, $\tau_1$, is obtained by
integrating $d\tau_1 = \sigma_T n_e(z) dz$, from $z=z_b$ to infinity,
where $n_e(z)$ is the electron density; $z$ is the vertical
coordinate; $ z_b\simeq H$ gives the location of the bottom of the
ionized skin; and $H$ is the disk scale height. Note that a simpler
estimate of $\tau_1$ can be obtained if one assumes that the gas
temperature is equal to the Compton one, and that the density law
follows a Gaussian law (see Kallman \& White 1989).

In our calculations of $\tau_1$, we will neglect by the reprocessing
of the radiation as it penetrates through the skin.  This limits the
applicability of our results to $\tau_1\simlt 1$ if the angle,
$\theta$, that the X-ray radiation makes with the normal to the disk,
is not too large, i.e., that $\mu_i \equiv \cos\theta$ is not too
small.  In the case when $\mu_i\ll 1$, the spectral reprocessing
cannot be neglected for $\tau_1\simgt \mu_i$. For simplicity, we will
assume $\mu_i\simlt 1$ and postpone the treatment of large incident
angles to a future publication.

If $\pmin$ is the gas pressure at $z=z_b$, then we can define
dimensionless gas pressure as $\pstar \equiv P/\pmin$, where
$P=\rho/\mu kT$ and $\rho$ is the gas density, and also dimensionless
gas temperature by $\tstar \equiv T/T_c$ ($T_c$ is the Compton
temperature, see below). The Compton-heated skin then has $\pstar <
1$, whereas the colder material below the skin has $\pstar > 1$.  As
shown in paper I, the incident X-rays do not affect the hydrostatic
balance in the ionized skin because the main source of opacity is
Compton scattering (see Fig. 5b in paper I and also Sincell \& Krolik
1997). With this, one can re-write hydrostatic balance equation in
terms of dimensionless variables as
\begin{equation}
{\partial P_*\over \partial x} = - 2 {P_*\over \tstar}\; {x -
\zeta\over \lambda^2}\;, 
\label{heq1}
\end{equation}
where $x\equiv z/H$, $\zeta$ is the ratio of the disk {\it midplane}
radiation pressure to the total pressure $\zeta =
\prad(0)/(\prad(0)+\pg(0))$, and $\lambda$ is the scale-height of the
skin in units of $H$:
$\lambda^2 = (2 kT_c R^3/ GM \mu H^2)$.
Expressing the electron  density as $n_e = \rho/\mu_e$, one can
arrive at
\begin{equation}
\tau_1 \,=\, {\mu\over \mu_e}\; {c\pmin\over \fx}\; {l_x\over
\theta_c}\;\lambda\,\int_{y_b}^{\infty} \;{\pstar\over\tstar} \;
dy\equiv \tau_0\; W(y_b)\;,
\label{first}
\end{equation}
where $l_x$ is the compactness parameter of the illuminating X-rays
(see, e.g., equation 20 in paper I); $\theta_c \equiv kT_c/m_ec^2$; we
also defined $y \equiv (x-\zeta)/\lambda$ and $y_b = (z_b/H
-\zeta)/\lambda$; finally, $\tau_0$ is the expression before the sign
of the integral. The integral in the equation (\ref{first}) is
designated as $W(y_b)$ and often (see below) turns out to be of order
unity. Krolik, McKee \& Tarter (1981, \S IV2b; KMT hereafter) showed
that for a completely ionized gas the energy balance equation can be
written as
\begin{equation}
\tstar^2 + \tstar^{1/2}\,\Xi_*^{-1} - \tstar = 0\;,
\label{en1}
\end{equation}
where $\Xi_*\equiv \Xi/\Xi_{ic}$ is the pressure ionization parameter
normalized by the ``inverse Compton'' ionization parameter $\Xi_{ic}$,
which is given by equation (4.5) of KMT, re-written in order to take
into account the difference in definitions of $\Xi = \fx/cP$ (the one
used here) and $\Xi = F_{ion}/n_H kT$ (KMT; $F_{ion}$ is the X-ray
flux between 1 and 10$^3$ Ry): $\Xi_{ic} = 0.47 * T_8^{-3/2}$; $T_8$
here is the Compton temperature in units of $10^8$ Kelvin.

The solutions presented in paper I possess the property that the
transition from the Compton-heated to cooler layers occurs at the
upper bend of the S-curve (e.g., point (c) in Fig. 1 of paper I). In
that point, $dT/d\Xi = \infty$. Using this condition, one can show
that the transition happens at $\tstar = 1/3$ where $\Xi =
{3^{3/2}\over 2}\, \;\Xi_{ic}$. Therefore, 
\begin{eqnarray}
\pmin = {2\over \,3\sqrt{3}}\; \Xi_{ic}^{-1} \; {\fx\over c}\;
= 3.2\times 10^{-2} T_1^{3/2}\;{\fx\over c}\;
\label{pmin}\\
\tau_1 \simeq 6.9 \; T_1\, {\fx\over F_d}\, G^{1/2}(r)\, \dm \, (1-f)
\label{tauh}\\
G(r) \equiv \,{2^{16}\over 27}\; [1- (3/r)^{1/2}]^2\, r^{-3}
\;,\nonumber
\end{eqnarray}
where $T_1 = kT_c/1$ keV; $r= R/R_s$, $R_s = 2GM/c^2$ is the
Schwarzschild radius; $\dm$ is the dimensionless accretion rate ($\dm =
1$ corresponds to Eddington luminosity for the accretion disk) and
$0\leq f < 1$ is the coronal dissipation parameter (e.g., Svensson \&
Zdziarski 1994). The maximum of function $G(r)$ occurs at $r = 16/3$
where it is equal to unity. We also approximated $\mu_e = m_p$, $\mu =
m_p/2$, and $W(y_b)\simeq W(0) = 1.23$ for the following reason. If
the vertical pressure profile is given by $P(z) = P(0) \exp[-(z/H)^2]$
for the gas-dominated disk, the location of the temperature
discontinuity is
\begin{equation}
z_b = H \;\ln^{1/2}\left[ {P(0)\over \pmin}\right]\;
\label{zb}
\end{equation}
The function $\ln^{1/2}(t)$ is a very slow, monotonically increasing
one: for example, its value is $1.52$ for $t = 10$ and $3.72$ for
$t=10^6$.  Therefore, for most realistic situations, $z_b$ is not very
much larger than $H$. As is easy to check, $\lambda \gg 1$ for
gas-dominated disks, and thus $y_b\ll 1$. For radiation-pressure
dominated disks, $z_b$ is nearly equal to $H$ (see \S 3.4 in paper I),
and hence $y_b = (1 - \zeta)/\lambda \ll 1$ in this case as well.  To
summarize this statement in words, the vertical extent of the ionized
skin is always large enough that the exact location of the inner
boundary is unimportant, for either gas or radiation-dominated disks.

A cautionary note is in place. So far we assumed that the spectrum of
the ionizing radiation does not change with the depth into the
illuminated material. This is true only for optically thin situations,
i.e., when $\tau_1\ll 1$, but for larger optical depths the Compton
temperature is smaller than the one corresponding to $\tau_1=0$. The
decrease in the local value of $T_1$ dictates a corresponding decrease
in the value of $\pmin$ according to equation (\ref{pmin}), and
therefore the transition to the cold equilibrium solution will always
happen at a smaller $\tau_1$ than equation (\ref{tauh}) predicts.  To
reflect this fact, Kallman \& White (1989) introduced $\tau_{\rm Th \
crit}\sim 1-10$, defined to be the maximum of $\tau_1$. Note that the
behavior of $\thot$ with Compton temperature, X-ray flux and radius
given by equation (\ref{tauh}) coincides with that obtained by these
authors ($T_{\rm IC8}$ on the second line of their equation [1] should
be in power $+1$ rather than $-1/2$ [T. Kallman 1999, private
communication]).

With the understanding that for $\tau_1\simgt 1$ our analysis may not
be expected to be very accurate, we {\em choose} to fix $\tau_{\rm Th\
crit} $ at a value of 3. This number is motivated only by the
convenience -- in paper I, due to the large volume of calculations, we
have limited the Thomson depth of the illuminated gas that can be
strongly ionized to $\tau_T = 3$. This choice did not lead to any
spurious results because the reprocessing features (from the cold
material below the skin) become negligible already at $\tau_T =2$ (see
paper I). In this paper, we use the following simple modification to
the Thomson depth of the hot skin:
\begin{equation}
\thot = {\tau_1\over 1 + \tau_1/3}\;,
\label{tauh1}
\end{equation}

\begin{figure*}[T]
\centerline{\psfig{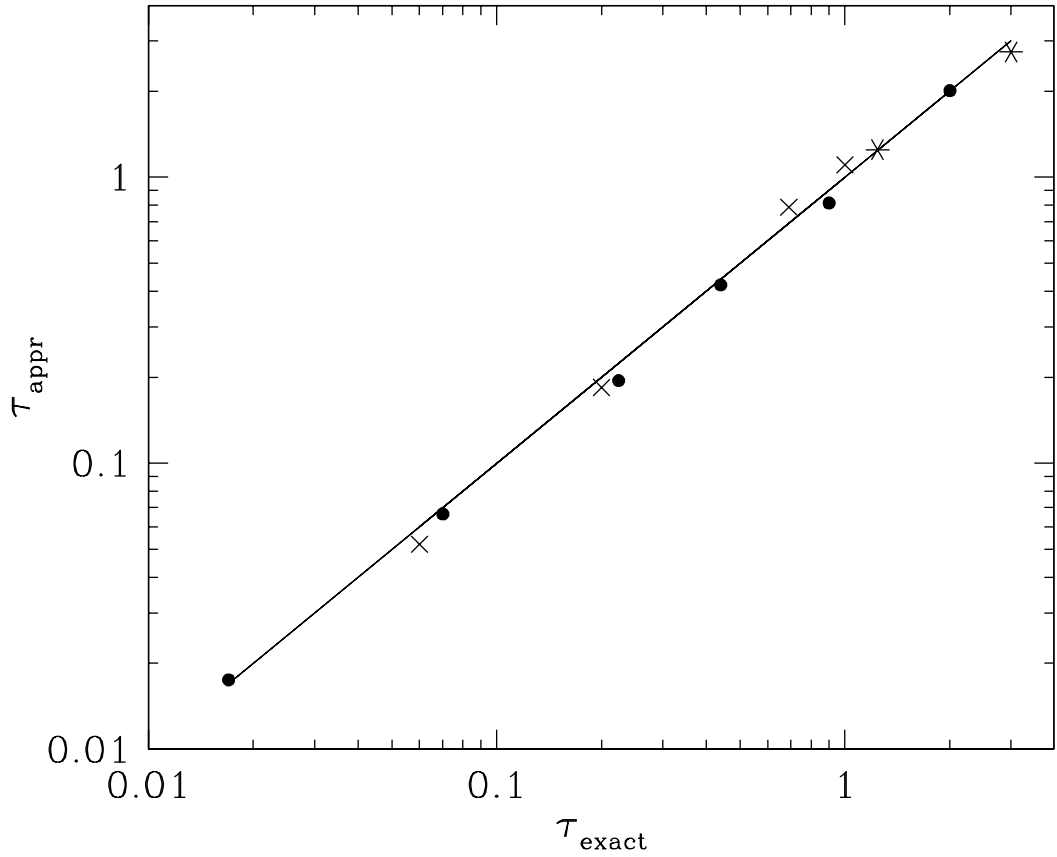}}
\caption{Comparison of the analytical approximation to the Thomson
depth of the hot skin and the numerical results presented in paper I.
Filled circles show tests with $\Gamma = 1.9$, but with varying
gravity parameter $A$ (\S\S 4.2-4.4 in paper I); crosses show tests
with $A=0.3$ and with $\Gamma$ varying between 1.5 and 2.3 (\S 4.7 in
paper I); and stars show the radiation-dominated case (described in \S
5 -- dotted and long-dashed curves in Fig. 13 in paper I).  The solid
curve shows the line $\tau_{\rm approx}= \tau_{\rm exact}$.}
\end{figure*}

We can now compare our results with values of $\thot$ numerically
calculated in paper I. There, we conducted two types of tests. For
radiation-dominated disks, we used fully self-consistent formalism and
thus we can directly compare results of \S 5 in paper I to results of
equations (\ref{tauh}) and (\ref{tauh1}). For purposes of isolating
dependence of the reflected spectra on one parameter at a time and yet
covering a large parameter space, we also conducted several tests
where we artificially varied the ``gravity parameter $A$'' (see \S 4
in paper I). In order to compare results of these latter tests, one
should use the same approach that led us to equation (\ref{tauh}), and
also use the same gravity law as used in paper I with (note that in
our current formulation, $A =4\theta_c\,l_x^{-1}\,\lambda^{-2}$).
This yields $\tau_1 = 0.73 \, T_1\, (l_x/A)^{1/2}$. Further, since
$\lambda$ is now expressed as $\lambda^2 = 4\theta_c\,
A^{-1}\,l_x^{-1}\,$, the value of $y_b$ is not necessarily small, and
thus $W(y_b)$ needs to be evaluated exactly.  Finally, to locate the
position of the lower boundary of the illuminated layer (which occurs
almost at the same height as the temperature discontinuity -- see
Fig. 5d in paper I for an example), we must substitute $\pmin$ in
equation (\ref{zb}) on the value of gas pressure at Thomson depth
$\tau_{\rm max}=4$, as used in paper I, which is approximately equal
to $\tau_{\rm max}\, A \,\fx/c = 4 A \fx/c$.

The results of such a comparison are summarized in Figure
(1), where we show $\thot$ obtained in this paper versus
that obtained numerically in paper I. The value of the Compton
temperature (i.e., $T_1$ in eq. \ref{tauh}) is taken to be the
temperature of the first zone in the temperature profiles shown in the
corresponding Figures in paper I. We note that the deviation of our
approximate expression from ``exact'' results is less than $\sim 20$\%
even though we covered a wide range of physical conditions, i.e.,
strong to weak illumination limits, different indices of the incident
X-ray radiation, and the disk itself is either gas or radiation
dominated.

\section{``Lamp Post'' Model}\label{sect:lamp}

As an application of our methods, we choose to analyze the model where
the X-ray source is located at some height $h_x$ above the black hole
(the ``lamp post model'' hereafter).  Iwasawa et al. (1996) reported
observations of iron line variability in Seyfert Galaxy MCG-6-30-15,
and pointed out several problems connected with theoretical
interpretation of these observations. Reynolds \& Begelman (1999)
showed that the accretion flow within the innermost stable radius may
be optically thick and thus produce fluorescent iron line emission in
addition to such emission from the disk itself. These authors argued
that the line emission from within the innermost stable radius may be
important for the interpretations of the observations of Iwasawa et
al. (see also Dabrowski et al. 1997).  Reynolds et al. (1999) studied
response of the iron line profiles to changes in the X-ray flux (iron
line reverberation).  In this paper we will not discuss the region
within the innermost stable orbit for the reason that the properties
of the accretion flow there are not well constrained.  In particular,
the hydrostatic and energy balance equations do not necessarily apply
since the hydrostatic and thermal time scales can be longer than the
in-fall time.

We will assume a non-rotating black hole and that all the X-rays are
produced within the central source. We also neglect all relativistic
effects in the present treatment. The projected X-ray flux impinging
on the disk is
\begin{equation}
\fx = {L_x h_x\over 4\pi \,(R^2 + h_x^2)^{3/2}} \;.
\label{fxscale}
\end{equation}
Let us define $\eta_x$ as the ratio of the total X-ray luminosity to
the integrated disk luminosity of the source, i.e., $\eta_x\equiv
L_x/L_d$. Using equation (\ref{tauh}), one obtains
\begin{equation}
\tau_1 \simeq 27.2 \; {h_x\over R_s}\, \eta_x\, T_1 \; [1 +
(h_x/R)^2]^{-3/2}\; r^{-3/2}\; \dm \;
\label{tau1}
\end{equation}
(one still has to use equation [\ref{tauh1}] to get the final answer
for $\thot$). For illustration, we choose a value of $h_x = 6 R_s$.
For $\Gamma = 1.9$, a typical value for Seyfert Galaxies, the Compton
temperature $kT_x \simeq 7.6$ keV. As discussed in paper I (\S 4.6),
the Compton temperature at the surface of the disk depends on the
cosine of the X-ray incidence angle, $\mu_i$, and the ratio
$\fx/\fdisk$ approximately as:
\begin{equation}
T_c \simeq T_x \left [ 1 + \mu_i \,\sqrt{3}\; {\fx + \fdisk
\over \fx}\right ]^{-1}\;.
\label{tx}
\end{equation}

Figure (2) shows the Thomson depth of the skin as a function of radius
for several values of $\dm$. The parameters in Figure (2) are chosen
to be: (a) $\Gamma =1.9$, $\eta_x = 1$ (``X-ray strong'' case); (b)
$\Gamma =1.9$, $\eta_x = 0.1$ (``X-ray weak'' case), and (c) $\Gamma =
2.3$, $\eta_x = 1$.  Figure (3) shows the angle-averaged reflected
spectra and temperature profiles of the hot layer for the three cases
just considered and the accretion rate equal to the Eddington value
($\dm = 1$) for $r=10$. The details of numerical methods with which
these spectra were obtained are described in paper I.

\begin{figure*}[T]
\centerline{\psfig{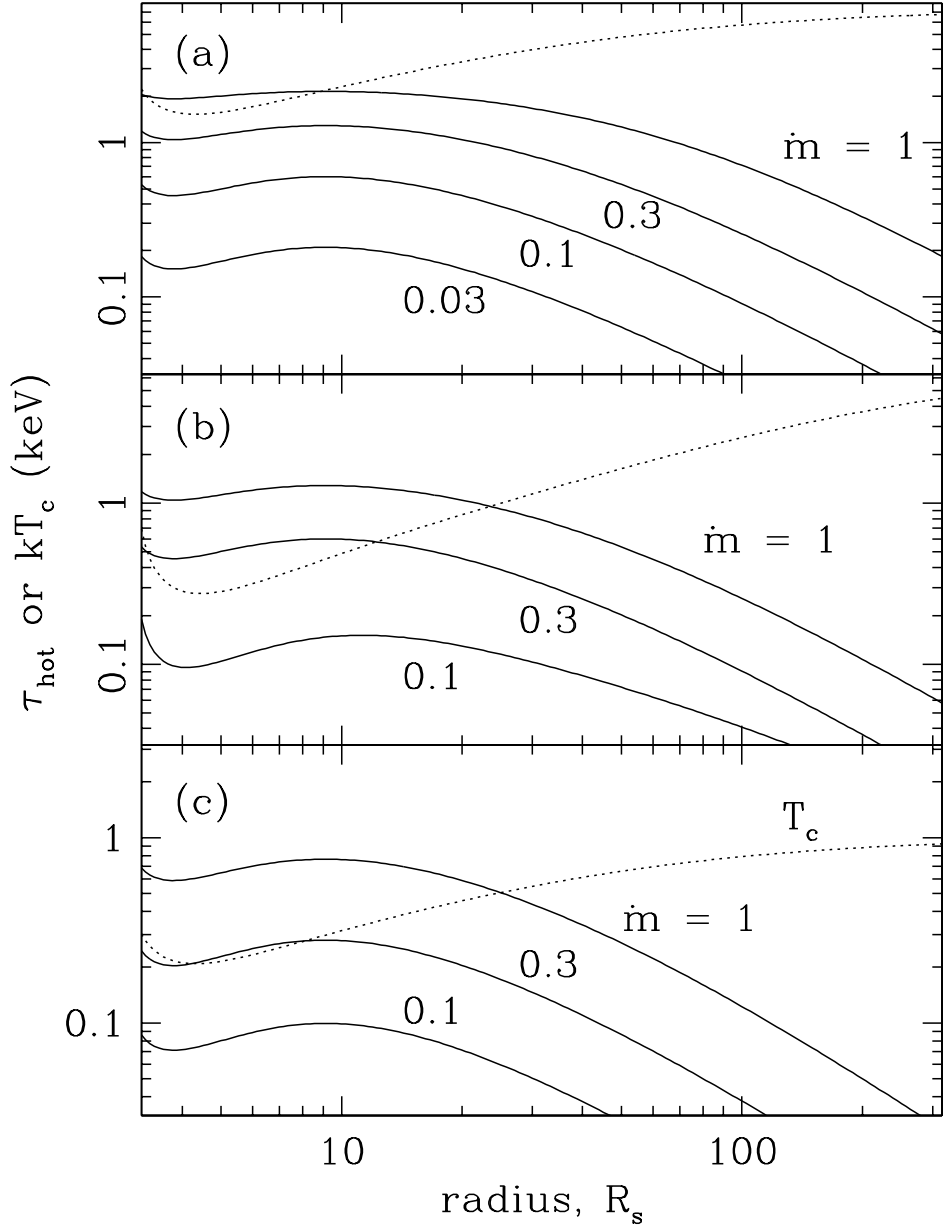}}
\caption{Thomson depth (solid curves) of the hot skin as a function
of radius for different accretion rates (whose values are shown next
to corresponding curves). The Compton temperature is shown by the
dotted curve. See text for additional details.}
\end{figure*}

If Compton temperature $T_c \simgt 1$, the illuminated gas is nearly
completely ionized, and the line is emitted almost exclusively from
the material below the skin. For this reason, it turns out that local
emissivity of the line is negligible when $\thot\simgt 1$ (see paper
I). Thus, the strength of the iron line will be decreasing with $\dm$
for the X-ray strong case (Fig. 2a) when $L_x\simgt$ few percent of
the Eddington value. The skin is thickest for smaller radii, and
therefore the broad iron line component will decrease first. The
narrow line component (emitted farther away from the black hole) will
also decrease with X-ray luminosity, but considerably slower than the
broad line. In addition, if some of the line comes from a putative
distant obscuring torus (Krolik, Madau \& \zycki 1994), or the disk
has a concave geometry (Blackman 1999), then the narrow component will
decrease even slower.  Thus, our qualitative description of the
behavior of the iron line EW with luminosity for X-ray strong sources
is consistent with Figure (3) of Nandra et al. (1997b), suggesting an
explanation to the X-ray Baldwin effect.

The X-ray weak case (Figs. 2b \& 3b) is different in two
respects. Firstly, the Thomson depth of the hot skin is smaller at a
given accretion rate compared with the X-ray strong case. Most
importantly, however, the skin is not ``that hot''. Namely, the skin
temperature is only $\sim 0.3$ keV in the inner disk, and some of the
iron ions are not completely ionized. For that reason,
it turns out that the iron line centroid energy is close to 6.7 keV,
and a very deep absorption edge appears. This edge is in fact much
stronger than the one resulting from a neutral material. Hence it is
possible that this relatively cold X-ray heated skin can be
unambiguously detected in spectra of real AGN.

\begin{figure*}[T]
\centerline{\psfig{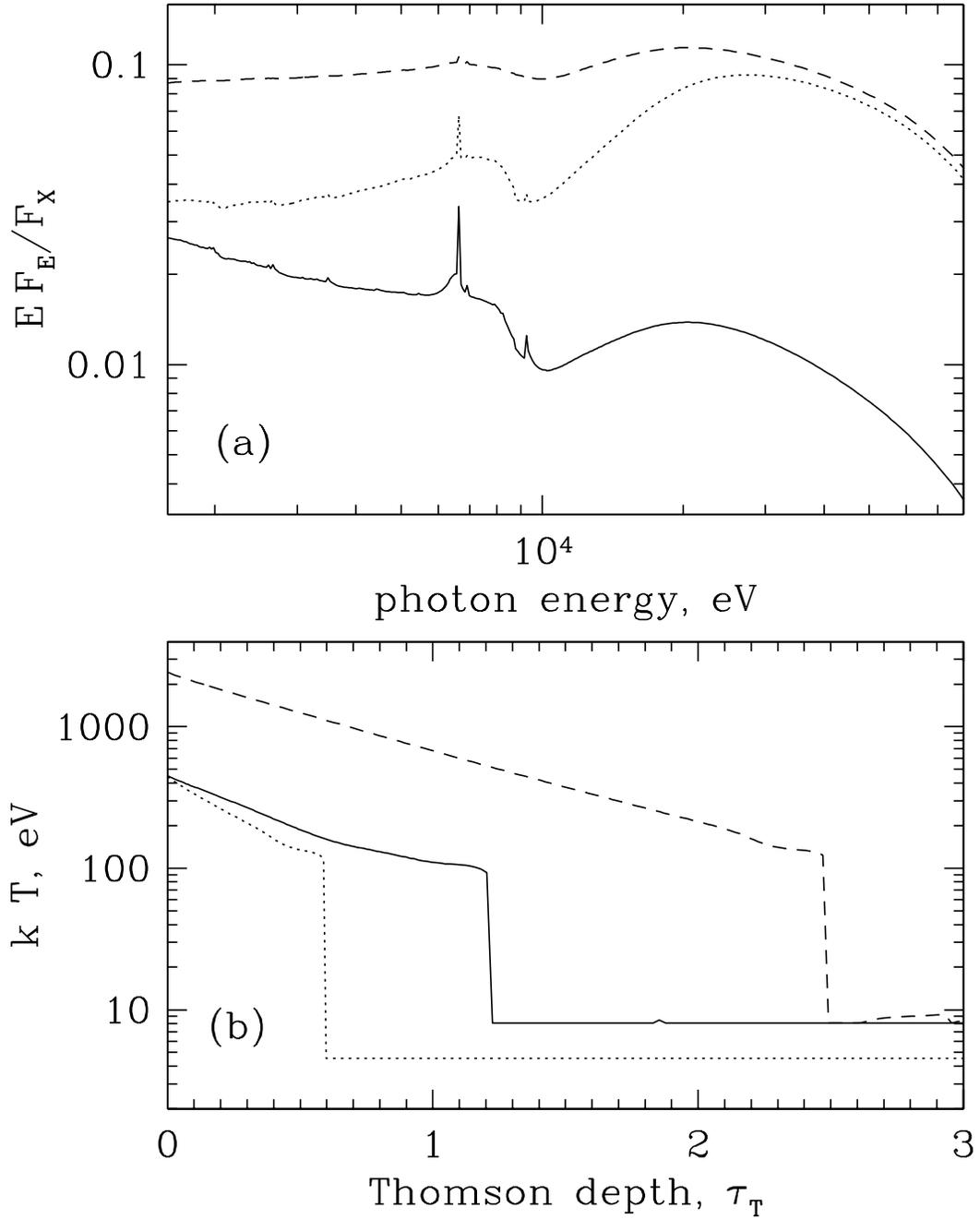}}
\caption{Reflected spectra (a) and temperature profile (b) computed
for the lamp post geometry at $r=10$ with accretion rate $\dm=1$. The
dashed, dotted and solid curves correspond to models shown in Fig. (2)
[a], [b] and [c], correspondingly. The solid curve in [a] was scaled
down by factor of 2 for clarity.}
\end{figure*}

Similarly, the soft incident X-ray spectrum leads to a relatively cool
skin because $T_x\sim 1$ keV only (see Fig. 2c \& 3c). As in the case
$\eta_x=0.1$, a strong absorption edge is observed. In addition, the
6.7 keV Fe line is stronger, with EW of 65 eV. Therefore, such skin
may also be detectable if it exists.

Comparing the value of $\thot$ resulting from analytical formulae with
those seen in Fig. (3b), one notes that deviations are as large as
$\sim$ 50\%. These relatively large deviations exemplify the fact that
our equations (\ref{tauh}) and (\ref{tauh1}) are good approximations
only to cases with strong X-ray flux and hard incident spectra (i.e.,
$\fx\simgt F_d$ and $\Gamma\simlt 2$). When one of these two
conditions is not satisfied (and it is the case for all three cases
presented in Figs. 2 \& 3) the Compton heated layer is cool enough for
atomic processes to provide additional sources of heating and cooling
beyond Compton and bremsstrahlung processes, so that the energy
equation (\ref{en1}) is not obeyed. However, our results can still be
used as an order of magnitude estimate of $\thot$ in the latter cases.

\section{Summary}\label{sect:summary}

In this Letter, we derived an approximate expression for the Thomson
depth of the hot completely ionized skin on the top of an accretion
disk illuminated by X-rays. Our results are only weakly dependent on
the a priori unknown $\alpha$-viscosity parameter (because it enters
through the boundary conditions). This allows us to reduce the
uncertainty in documenting predictions of different accretion theories
with respect to the iron line profiles and the strength of the X-ray
reflection hump. Using the ``lamp post model'' geometry as an example,
we showed that an inner part of an accretion disk may have a
Thomson-thick ionized skin. Under certain conditions ($\fx\simgt F_d$
and $\Gamma\simlt 2$), the X-ray heated skin may act as a perfect
mirror for photons with energies below $\sim 30$ keV. The physical
cause of this is that Compton scattering in the skin prevents X-rays
from penetrating to the deep cold layers that are capable of producing
fluorescent line emission as well as setting other marks of atomic
physics.

We note that because the ionized skin is thickest in the inner disk,
the observed absence or deficit of the relativistically broadened line
and other reprocessing features in some systems (e.g., \zycki, Done \&
Smith 1997, 1998), which was interpreted as a possible evidence for a
disruption of the cold disk for small radii, may also mean that the
``cold'' disk is still present up to the innermost radius, but the
skin effectively shields it from the X-ray illuminating flux. It is
interesting, however, that the presence of the skin becomes apparent
in systems that have $\fx\simlt F_d$ or $\Gamma\simgt 2$, because the
Fe atoms are not completely stripped of their electrons and thus
produce strong ionized edges and lines. We believe these predictions
should be testable observationally with current X-ray missions such as
Chandra, Astro-E and XMM. Also note that for a patchy corona model of
accretion disks (e.g., Haardt, Maraschi \& Ghisellini 1994), the
Thomson depth of the hot skin is always larger than the one found
here, since the ratio $\fx/F_d$ is larger.  Finally, the presence of
the ionized skin is important not only for the X-rays, but for other
wavelengths as well (e.g., in studies of Lyman edge of accretion
disks, correlation of optical/UV light curves with X-rays).

A shortcoming of this paper is that we neglected the changes in the
ionizing continuum due to scattering in the skin, which, strictly
speaking, limits the applicability of our results to optically thin
situations $\thot\simlt 1$ and incident angles not too far from
normal\footnote{A crude ``fix'' to the problem of small incident
angles is to use $c I_x$, where $I_x$ is the X-ray intensity
integrated over $\phi$ angle, in place of $\fx$ in equation (5)}. We
plan to address this issue in future.

The author acknowledges support from NAS/NRC Associateship and many
useful discussions with D. Kazanas, T. Kallman and M. Bautista.  The
anonymous referee is thanked for several constructive suggestions.

{}

\end{document}